\documentclass[superscriptaddress,letterpaper, amssymb,preprint, amsmath, 11pt]{revtex4-1}
\usepackage[english]{babel}
\usepackage[utf8]{inputenc}
\usepackage{mathtools}
\usepackage{braket}
\usepackage{xcolor}
\usepackage[colorlinks,citecolor=red,urlcolor=blue,bookmarks=false,hypertexnames=true]{hyperref} 
\usepackage{caption,subcaption}
\usepackage[justification=raggedright,singlelinecheck=false,labelfont=bf]{caption}
\usepackage{siunitx}

\usepackage{scalerel}
\usepackage{tikz}
\usetikzlibrary{svg.path}
\definecolor{orcidlogocol}{HTML}{A6CE39}
\tikzset{
  orcidlogo/.pic={
    \fill[orcidlogocol] svg{M256,128c0,70.7-57.3,128-128,128C57.3,256,0,198.7,0,128C0,57.3,57.3,0,128,0C198.7,0,256,57.3,256,128z};
    \fill[white] svg{M86.3,186.2H70.9V79.1h15.4v48.4V186.2z}
                 svg{M108.9,79.1h41.6c39.6,0,57,28.3,57,53.6c0,27.5-21.5,53.6-56.8,53.6h-41.8V79.1z M124.3,172.4h24.5c34.9,0,42.9-26.5,42.9-39.7c0-21.5-13.7-39.7-43.7-39.7h-23.7V172.4z}
                 svg{M88.7,56.8c0,5.5-4.5,10.1-10.1,10.1c-5.6,0-10.1-4.6-10.1-10.1c0-5.6,4.5-10.1,10.1-10.1C84.2,46.7,88.7,51.3,88.7,56.8z};}}
\newcommand\orcidicon[1]{\href{https://orcid.org/#1}{\mbox{\scalerel*{
\begin{tikzpicture}[yscale=-1,transform shape]
\pic{orcidlogo};
\end{tikzpicture}
}{|}}}}

\begin{document}
\title{Characterization of self-heating in cryogenic high electron mobility transistors  using Schottky thermometry}

\author{Alexander Y. Choi \orcidicon{0000-0003-2006-168X}}
\affiliation{Division of Engineering and Applied Science, California Institute of Technology, Pasadena, CA 91125, USA}

\author{Iretomiwa Esho \orcidicon{0000-0002-3746-6571}}
\affiliation{Division of Engineering and Applied Science, California Institute of Technology, Pasadena, CA 91125, USA}

\author{Bekari Gabritchidze \orcidicon{0000-0001-6392-0523}}
\affiliation{Division of Physics, Mathematics, and Astronomy, California Institute of Technology, Pasadena, CA 91125, USA}

\author{Jacob Kooi \orcidicon{0000-0002-6610-0384}}
\affiliation{NASA Jet Propulsion Laboratory, California Institute of Technology, Pasadena, CA 91109, USA}

\author{Austin J. Minnich \orcidicon{0000-0002-9671-9540}}
\thanks{Corresponding author: \href{mailto:aminnich@caltech.edu}{aminnich@caltech.edu}}
\affiliation{Division of Engineering and Applied Science, California Institute of Technology, Pasadena, CA 91125, USA}

\date{\today} 

\begin{abstract}
Cryogenic low noise amplifiers based on high electron mobility transistors (HEMTs) are widely used in applications such as radio astronomy, deep space communications, and quantum computing, and the physical mechanisms governing the microwave noise figure are therefore of  practical interest. In particular, the contribution of thermal noise from the gate at cryogenic temperatures remains unclear owing to a lack of experimental measurements of thermal resistance under these conditions. Here, we report measurements of gate junction temperature and thermal resistance in a HEMT at cryogenic and room temperatures using a Schottky thermometry method. At  temperatures $\sim 20$ K, we observe a nonlinear trend of thermal resistance versus power that is consistent with heat dissipation by phonon radiation. Based on this finding, we consider heat transport by phonon radiation at the low-noise bias and liquid helium temperatures and estimate that the thermal noise from the gate is several times larger than previously assumed  owing to self-heating.  We conclude that without improvements in thermal management, self-heating results in a practical lower limit for  microwave noise figure of HEMTs at cryogenic temperatures.
\end{abstract}
\maketitle

\section{Introduction}

Microwave low noise amplifiers (LNAs) based on high electron mobility transistors (HEMTs) are widely-used components of scientific instrumentation in fields such as  radio astronomy \cite{Bryerton_2013,pospi_2005}, deep space communication \cite{bautista_2001}, and quantum computing \cite{krantz_2019,Hornibrook_2015,Chow_2014,Cha_2020,Arute2019}.  After decades of development \cite{Cha_2018,Heinz_2020,Akgiray_2013,Varonen_2013,Cuadrado_2017}, HEMT LNAs have achieved cryogenic noise temperatures approximately 5-10 times the quantum limit over frequencies from 1-100 GHz \cite{Bryerton_2013}. Despite this progress, applications drive the development of amplifiers with ever-lower noise figures.

Noise in HEMT amplifiers is typically interpreted using the Pospieszalski model \cite{pospi_1989}. In this model, noise is decomposed into hot electron noise added in the channel and thermal noise in the gate, parameterized by equivalent temperatures $T_d$ and $T_g$, respectively. The gate noise temperature is typically assumed to be the cryostat base temperature $T_g = T$ while the drain temperature is fit to measured data. For a constant drain current, the hot electron contribution is taken to be constant and the minimum noise figure then scales as $T_g^{1/2}$ \cite{pospi_2005}.

Although the noise temperature does decrease with base temperature over a range of temperatures as predicted, at liquid helium temperatures the noise temperature is observed to plateau to a value several times the quantum noise limit \cite{schleeh_2014,Duh_1989,McCulloch_2017}. This noise temperature plateau has been attributed to a variety of mechanisms, including drain shot noise \cite{Pospieszalski_2017}, gate leakage current \cite{Cha_2018}, and self-heating \cite{Weinreb_1980, schleeh_2014}. In particular, Ref.~\cite{schleeh_2014} used measurements of microwave noise to conclude that the thermal resistance associated with phonon radiation leads to non-negligible self-heating at cryogenic temperatures. However, this conclusion is based on an indirect estimate of the gate junction temperature using a noise model.

Measurements of the gate temperature under bias at cryogenic temperatures would provide more direct evidence that self-heating is the origin of the noise temperature plateau. This measurement is challenging for conventional thermometery techniques such as IR microscopy \cite{Mittereder_2002, Shakouri_2009}, micro-Raman spectroscopy \cite{Kuball_2003,Choi_2013,Graham_2007}, or liquid crystal thermography \cite{Park_2003} due to geometrical constraints like the sub-micron gate lengths and the buried structure of modern HEMTs. Consequently, self-heating in FETs is usually characterized with measurements of temperature-sensitive electrical parameters. Early semi-quantitative studies of self-heating in CMOS estimated the temperature under bias using the temperature-dependence of drain current \cite{Takacs_1987, Craig_1972,Foty_1987,Foty_1989}. However, these approaches neglected a number of mechanisms relevant to the drain current  in sub-micron  devices such as the bias dependence of threshold voltage, series resistances, and non-stationary transport effects, which are known to be important in modern HEMTs and could affect the extracted temperature rise. Later studies of self-heating in MOSFETs incorporated some of these effects and reported measurements of temperature rise and thermal time constants \cite{Tenbroek_1996}. Recent work in SOI MOSFETs reported that the dominant thermal resistance is due to the buried oxide layer \cite{Triantopoulos_2018}. Self-heating studies in HEMTs have largely focused on GaN power FETs at room temperature, where device lifetime is limited by channel heating \cite{Garcia_2016}. Typically, the temperature rise is extracted from pulsed current measurements on the gate \cite{darwish_2008,Wu_2018}, but this technique is generally unsuitable for cryogenic HEMTs where the thermal time constants are on the same order as the pulse duration \cite{mottet2005zuverlassigkeitsstudien}. As a result,  self-heating in cryogenic III-V HEMTs remains poorly characterized.

Here, we present  measurements of the junction temperature and thermal resistance of the metallic gate in low-noise GaAs HEMTs using Schottky thermometry. At cryogenic temperatures, we observe a nonlinear trend of the thermal resistance on dissipated power that is consistent with heat transport by phonon radiation. We  analyze the implications of this finding at the low-noise bias in which the drain and gate are forward and reverse biased, respectively, using a radiative circuit model. The model predicts that the gate self-heats to a value comparable to the physical temperature of the drain, contradicting    the typical assumption that the gate is isothermal with the base temperature. Our study thus implies that without improvements to device thermal management to remove heat from the gate,  self-heating results in a practical lower limit for HEMT microwave noise figure at cryogenic temperatures.

\section{Experiment} \label{sec:methods}

We measured the temperature of the gate-barrier junction of a discrete HEMT under DC bias using the Schottky thermometry method introduced in Ref.~\cite{Narhi_2011}. In brief, the method exploits the temperature dependence of the electrical parameters of the Schottky junction to infer the temperature rise of the junction using DC $I\textnormal{-} V$ characteristics and microwave $S$ parameters. The current in a Schottky diode is given by:

\begin{equation}\label{eqn:SchottkyDiode}
    I = I_{S} \, \text{exp}\bigg( \frac{q \, (V-IR_{S})}{\eta \, k_{B} T_j}\bigg)    
\end{equation}

where $q$ is the elementary charge, $k_{B}$ is the Boltzmann constant, $R_S$ is the parasitic series resistance, $I_S$ is the saturation current, $\eta$ is the ideality factor, and $T_j$ is the intrinsic junction temperature. 

To obtain $T_j$ for a given bias, the temperature dependence of $I_S$ and $\eta$ is extracted from DC $I\textnormal{-} V$ characteristics by varying the cryostat base temperature at low bias where self-heating is negligible. Next, the small-signal resistance about a DC bias is determined from the microwave $S$ parameters. Equating expressions for the series resistance yields an equation including the measured Schottky and $S$ Parameters in which the only unknown is the junction temperature:

\begin{equation}\label{eqn:rootEqn}
    \frac{V}{I} - \frac{\eta(T_j)\,k_B T_j}{qI} \, \text{ln}\bigg(\frac{I}{I_S(T_j)}\bigg) = r_T - \frac{\eta(T_j)\,k_B T_j}{qI}
\end{equation}

Here, $r_T$ is the small signal resistance obtained from the S parameters. With the DC $I\textnormal{-}V$ characteristics and S parameters for different biases known, we obtain the junction temperature at various base temperatures by numerically solving Eqn.~\ref{eqn:rootEqn}.

We used this method to characterize the thermal resistance of a  70 nm gate length InAlAs/InGaAs mHEMT with a 200 \si{\micro \meter} gate width. Further details of the device are specified in Ch.~5.1 of Ref.~\cite{Akgiray_Thesis}. All measurements were performed in a custom cryogenic probe station \cite{Russell_2012} with cryostat base temperature (denoted base temperature, $T$) controlled between 20 K and 300 K by a LakeShore 336 temperature controller.  The HEMT was biased using a Minicircuits ZX85-12G-S+ bias-tee. $S$-parameter measurements were performed with a Rohde $\&$ Schwarz ZVA50 vector network analyzer from 10 MHz to 18 GHz, calibrated with the  through-reflect-match method. The DC measurements were corrected for loss in the bias-tee and coaxial lines. The temperature-dependence of the Schottky parameters $I_s$ and $\eta$ were extracted in the log-linear region at low bias $I \sim 100$ $\mu A$ so that self-heating can be neglected but at sufficiently high bias so that the characteristic is still log-linear \cite{Cibils_1985}. The saturation current and ideality factor were fit to the expressions derived from thermionic emission theory (See Eqns.~2 and 3 in Ref.~\cite{Narhi_2011}). The saturation current expression was modified by adding a temperature-independent term to Eqn.~3 in Ref.~\cite{Narhi_2011} to account for the leakage current that dominates at cryogenic temperatures. After extracting the junction temperature, we verified that the self-heating can be neglected in the fit range. The junction temperature was extracted at biases ranging from 0.5 to 10 \si{\milli \ampere} for which the capacitance of the channel depletion layer is negligible. We restrict the biases to this range to avoid non-negligible corrections associated with removing parasitic capacitance. For the calculation of the thermal resistance, we took the area of the gate to be 14 \si{\micro \squared \meter} $= 70$ \si{\nano \meter} $\times \, 200$ \si{\micro \meter}. 

\section{Results}
\begin{figure*} 
{
\includegraphics[width=\textwidth]{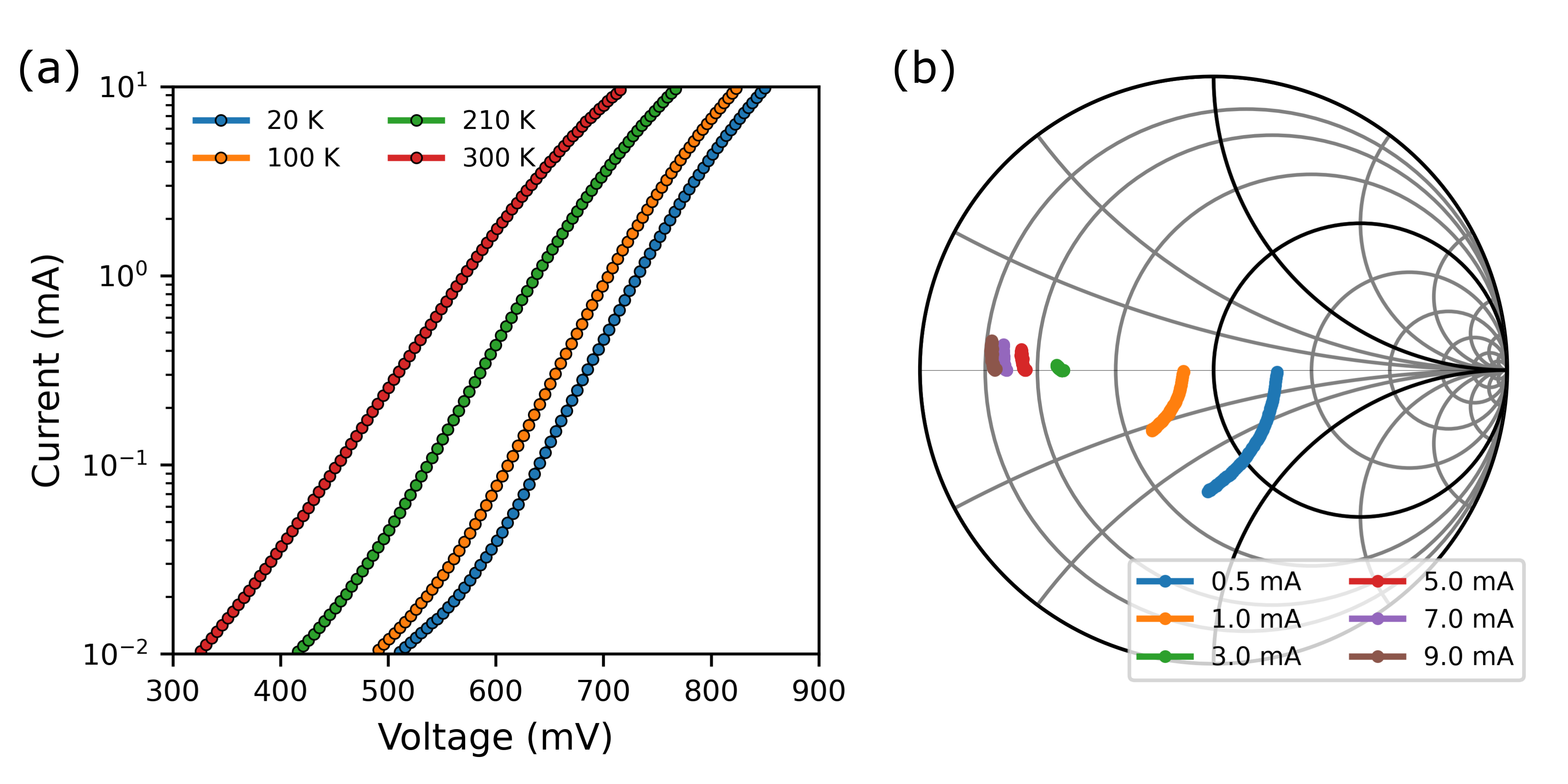}
\phantomsubcaption\label{fig:1a}
\phantomsubcaption\label{fig:1b}
}
\caption{
\textbf{(a)}~Measured forward $I\textnormal{-}V$ characteristics of the HEMT at different base temperatures (colored symbols). We fit the temperature-dependence of the Schottky parameters  for $I \lesssim 100$ \si{\micro \ampere} for which $T_j \approx T$ and the diode characteristics are log-linear. The DC resistance is combined with $S$ parameter measurements to extract the junction temperature at the 0.5-9 \si{\milli \ampere} biases for which self-heating occurs.
\textbf{(b)}~Small-signal reflection coefficient extracted from high frequency measurements at different bias points (colored markers) and $T=20$ K from 1-10 GHz. The small-signal resistance is extracted by taking the real part of the input impedance.}
\end{figure*}

\subsection{Schottky $I\textnormal{-}V$ measurements}

Figure~\ref{fig:1a} shows the measured forward bias DC $I\textnormal{-}V$ characteristics at different base temperatures. As the device is cooled, the characteristics shift to higher threshold voltages. The DC data were used to extract the temperature dependence of the saturation current and ideality factor as discussed in Section~ \ref{sec:methods}. Figure~\ref{fig:1b} shows the measured reflection coefficient $S_{11}$ at various bias points at 20 K from 1-10 GHz. The total small signal resistance is  obtained from the reflection coefficient through $r_T = \text{Re} [`  (1 + S_{11}) / (1 - S_{11})]$ where $Z_0$ is the characteristic impedance of the system. 

Qualitative evidence of self-heating can be obtained by  inspecting the $I\textnormal{-}V$ characteristics. Ideal Schottky diodes exhibit log-linear $I\textnormal{-}V$ characteristics, but in practice deviations are observed owing to  self-heating and parasitic series resistance. Series resistance leads to a sublinear log $I\textnormal{-}V$ while self-heating causes a superlinear trend (see Figure 3 in Ref.~\cite{Narhi_2011}). These effects depend on bias as well as  temperature, and their balance determines the trend of the measured $I\textnormal{-}V$ characteristic. 

\begin{figure*} [htb!]
{
\includegraphics[width=\textwidth]{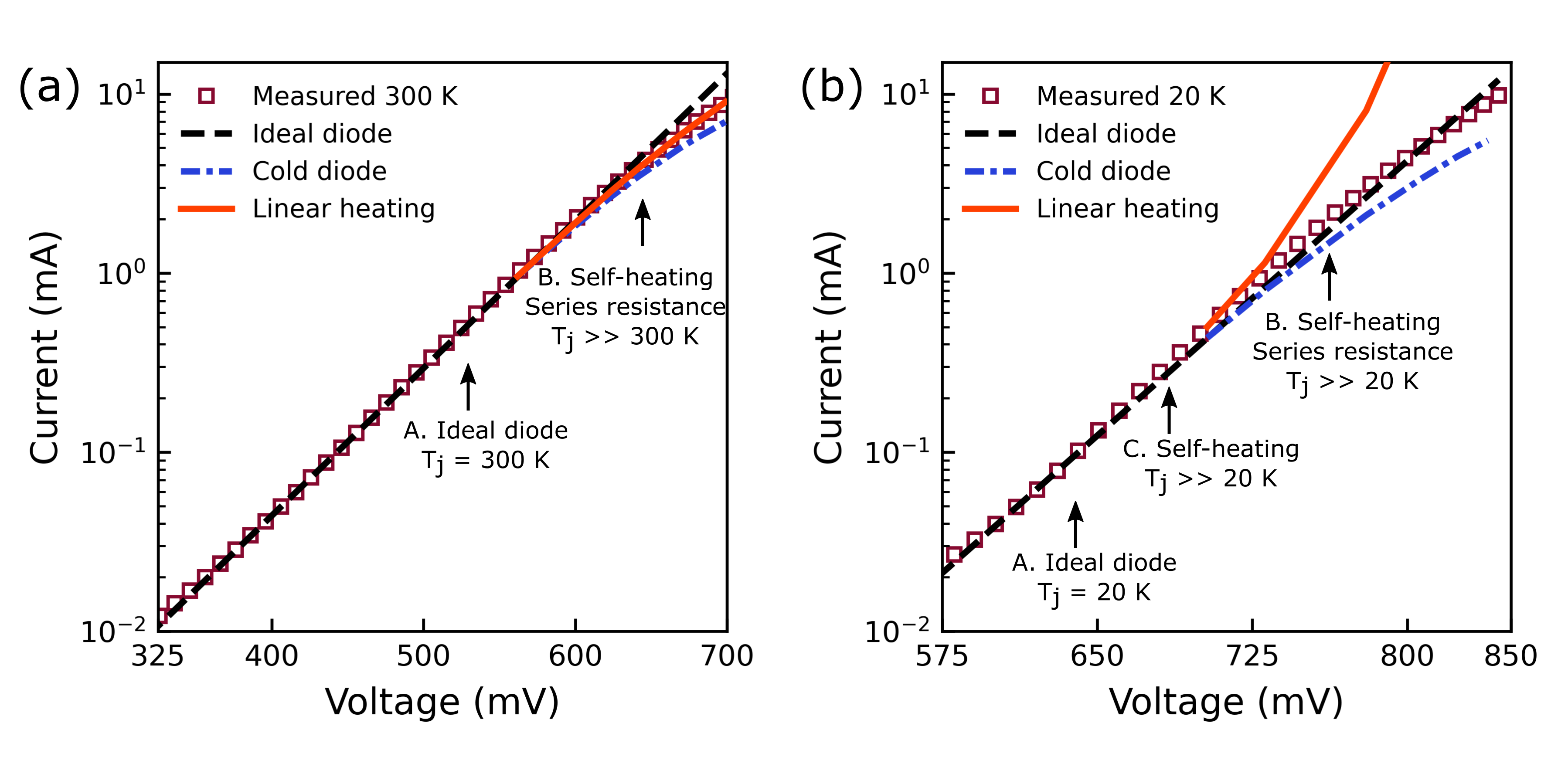}
\phantomsubcaption\label{fig:2a}
\phantomsubcaption\label{fig:2b}
}
\caption{Measured  $I\textnormal{-}V$ characteristics at \textbf{(a)}~300 K and
\textbf{(b)}~20 K (red markers) compared to the ideal diode (black dashed line), cold diode (blue dashed dotted line), and linear-heating model (orange solid line). A calculation assuming constant thermal resistance  explains the measurements at 300 K but not at 20 K. See text for detail. The curves for cold diode model and linear heating model coincide with the ideal diode model below 0.5 \si{\milli \ampere} and are omitted for clarity.}
\label{fig:2}

\end{figure*}

Figure~\ref{fig:2} compares the measured  $I\textnormal{-}V$ characteristics with those generated by an ideal Schottky model, a cold-diode model,  and a model including self-heating with a constant thermal resistance $R_{th}$.  The ideal Schottky model neglects  series resistance and self-heating and consequently exhibits log-linear DC characteristics at all biases. The cold-diode model incorporates the measured diode series resistance but neglects self-heating so that the junction temperature equals the base temperature at all biases, $T_j = T$. Finally, the linear-heating model incorporates series resistance, temperature-dependent Schottky parameters, and assumes that the junction temperature increases linearly with the dissipated power, $T_j =  T + R_{th} \, IV$, with thermal resistance $R_{th}$ taken as a fitting parameter. Through comparison to the measured DC $I\textnormal{-}V$, we can infer the relative magnitude and power dependence of the junction temperature $T_j$ at different $T$.

Figure~\ref{fig:2a} shows the model comparison to measurements at $T = 300$ K. At low biases below 1 \si{\milli \ampere} the measured diode is nearly ideal and exhibits the expected log-linear trend (Region A). At 1 mA, the cold-diode model, linear-heating model, and measured current agree to within 3\%, indicating that the temperature rise at this bias is small compared to the base temperature. At high currents exceeding 1 mA, the series resistance leads to a sublinear trend (Region B); however, the cold-diode model including only series resistance underpredicts the measured current at high biases ($\sim 9$ mA) by $\sim 25$\%. In contrast, the linear-heating model agrees with the measured $I\textnormal{-}V$ characteristics in Region B. From this comparison, we infer that at 300 K, self-heating is appreciable above 1 mA biases and that the thermal resistance is constant with power.

Figure~\ref{fig:2b} shows the model comparison at $T = 20$ K. As at 300 K, below 100 \si{\micro \ampere} the diode is nearly ideal, but a super-linear trend associated with self-heating is evident at intermediate biases between 100 \si{\micro \ampere} and 1 \si{\milli \ampere} (Region C), which is evident observed in the room temperature characteristics. At 1 mA, the measured current exceeds the cold-diode model by over 15\%, indicating that the temperature increase due to self-heating is substantially larger at this bias than at 300 K. Furthermore, above 1 mA (Region B) the linear-heating model at 20 K markedly overpredicts the measured current, indicating that the thermal resistance must decrease as the bias increases at 20 K.

\begin{figure*}[!htb]
{
\includegraphics[width=\textwidth]{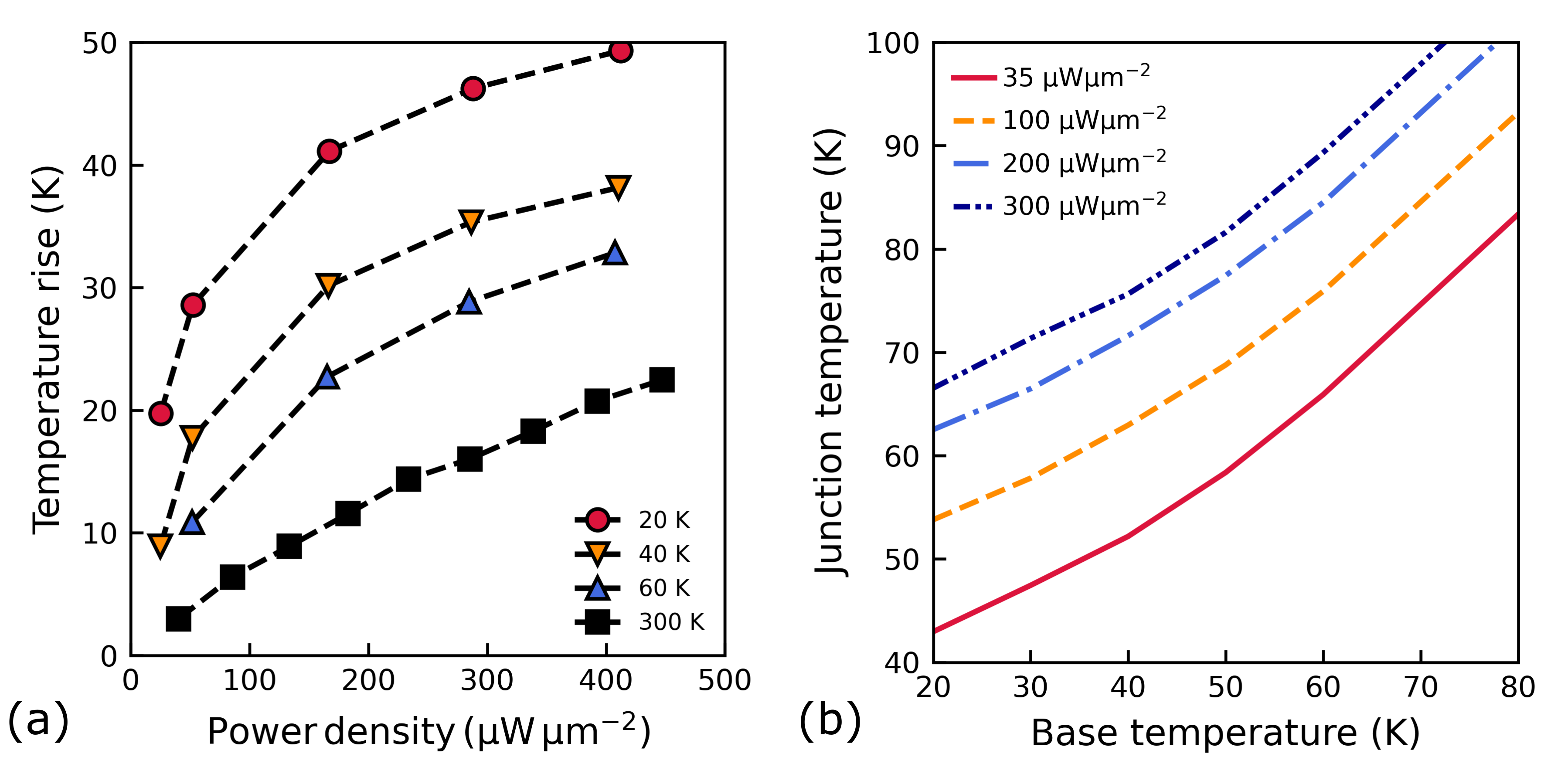}
\phantomsubcaption\label{fig:3a}
\phantomsubcaption\label{fig:3b}
}
\caption{\textbf{(a)}~Junction temperature rise, $T_j - T$, versus dissipated power density at base temperatures 20 K (red circles), 40 K (yellow triangles), 60 K (blue triangles), and 300 K (black squares). The temperature rise is approximately linear with power at room temperature but 
nonlinear at cryogenic temperatures. Dashed black lines are added as guides to the eye.
\textbf{(b)}~Interpolated junction temperature versus base temperatures at various power densities (colored lines). As the device is cooled, the junction temperature begins to plateau due to self-heating.}
\label{fig:3}
\end{figure*}

\subsection{Junction temperature extraction}

We now perform a quantitative analysis of the data by using the method in Sec.~\ref{sec:methods} to extract the junction temperature. Figure \ref{fig:3a} shows the extracted junction temperature rise versus power for $T =$ 300 K, 60 K, 40 K and 20 K. The features of the temperature rise are consistent with the qualitative expectations developed in Fig.~\ref{fig:2}. First, at $T = 300$ K, the temperature rise is nearly linear with the dissipated power, indicating that a constant thermal resistance can account for the measurements. Second, at the low bias point of 1 mA and 40 \si{\micro \watt \per \micro \meter \squared}, the junction temperature rise is 3 K or 1\% of the base temperature, confirming the qualitative prediction of small temperature rise at this power shown in Fig.~\ref{fig:2a}. In contrast, at cryogenic temperatures, the temperature rise exhibits a nonlinear trend with power, with the temperature initially increasing rapidly but transitioning to a weaker increase at higher powers. This observation is consistent with Fig.~\ref{fig:2b} and suggests that the thermal resistance decreases as the bias is increased. At $T = 20$ K and the same low bias point of 1 mA and 50 \si{\micro \watt \per \micro \meter \squared}, the temperature rise is 29 K, almost 10 times larger than the room temperature value of $ 3$ K. This difference is on the same order as the difference in heat capacity between these temperatures,  which decreases by over an order of magnitude from 300 K to 20 K \cite{Blakemore_1982}.
At $T=$ 40 K and 60 K, the temperature rise exhibits the same qualitative features as those seen at 20 K, but for the same power, the temperature rise is smaller at higher base temperatures.

As described in Sec.~\ref{sec:methods}, the temperature is extracted for fixed gate current values. To extract the temperature dependence of the junction temperature at fixed power, we linearly interpolate the junction temperatures as a function of power. This procedure is analogous to taking a vertical slice at fixed power in Fig.~\ref{fig:3a}. Figure~\ref{fig:3b} shows the interpolated junction temperature versus the base temperature for various power densities applied to the gate. At base temperatures near 80 K and all powers, the junction temperature decreases with base temperature as  $\Delta T_j/\Delta T \sim  0.9$ K/K, meaning that at these  temperatures, the junction and cryostat are cooled at nearly the same rate. As the cryostat is cooled to 20 K, the rate drops to $\sim$ 0.45 K/K, indicating that while the device continues to cool, the gate temperature cools less rapidly due to self-heating. The observed temperature saturation implies a nonlinear increase of the thermal resistance as the junction temperature approaches $\sim 50$ K. 

We now compute the thermal resistance as the ratio of the junction temperature rise in Fig.~\ref{fig:3a} and the power density. Figure~\ref{fig:4} shows the thermal resistance  versus power density. At room temperature, the thermal resistance is nearly constant with power, as expected from Figs.~\ref{fig:2a} and \ref{fig:3a}. As the device is cooled to cryogenic temperatures, the thermal resistance increases at all powers. At 1 mA and 40 \si{\micro \watt \per \micro \meter \squared}, the thermal resistance increases by almost an order of magnitude from 300 K to 20 K. Furthermore, at cryogenic temperatures the thermal resistance exhibits a nonlinear power dependence.

\subsection{Phonon radiation resistance}

We now examine the magnitude of the measured thermal resistance in context of values reported for various metal-semiconductor interfaces. The HEMT gate is formed by depositing a metallic stack consisting of metals such as Pt, Ti, and Au on the InAlAs barrier layer that has been subjected to  semiconductor processing steps such as wet etching. At 300 K, reported values of thermal boundary conductance for a soft metal such as gold on semiconductor are in the range of 30-40 \si{\mega \watt \per \square \meter \per \kelvin} \cite{Stoner_1993}. These studies utilize pristine interfaces for which the metal is evaporated directly onto a high-quality crystalline substrate. In contrast, the etching step in the fabrication of the gate leaves  an amorphous region several \si{\nano \meter} thick at the gate-semiconductor junction (see Figure 4.11(a) in Ref.~\cite{Cha_Thesis}). Prior measurements report that crystalline disorder can suppress the thermal conductance and enhance thermal boundary resistance by factors of approximately 3-4 (see Figure 7 in Ref.~\cite{Hopkins_2013}) as phonons with atomic-scale wavelengths are reflected at the interface \cite{Hua_2017}. At 300 K, the average thermal resistance of the HEMT over all power levels is 60 \si{\kelvin \square \micro \meter \per \milli \watt}. This value corresponds to a conductance of 17 \si{\mega \watt \per \square \meter \per \kelvin}, which is consistent with the above values for thermal conductance of a defective interface.

\begin{figure*}[!htb]
{
\includegraphics[width=\textwidth/2]{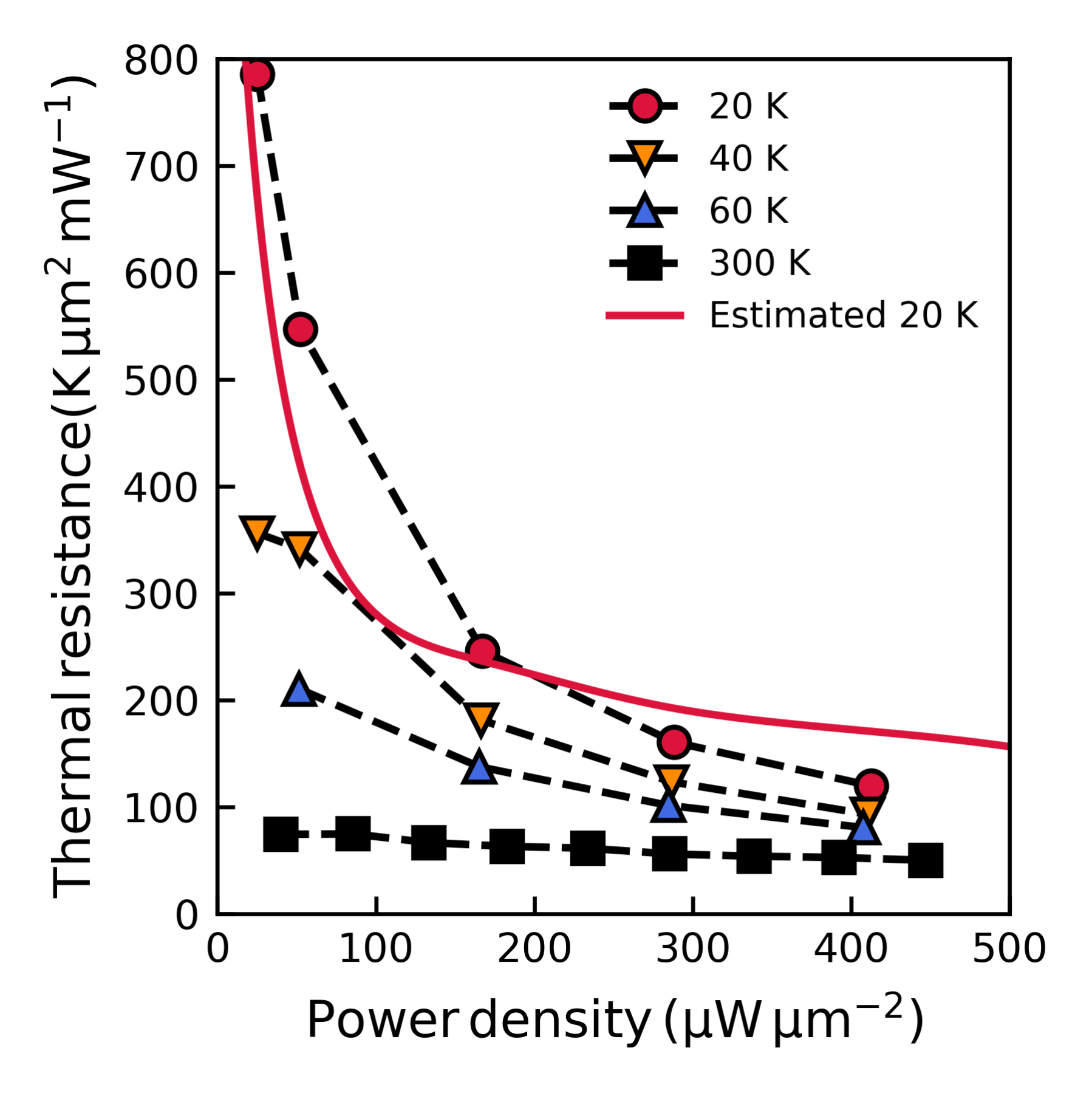}
}
\caption{Thermal resistance of the junction versus power density at base temperatures 20 K (red circles), 40 K (yellow triangles), 60 K (blue triangles), and 300 K (black squares). At room temperature, the thermal resistance is nearly independent of power and thus junction temperature. At cryogenic temperatures, the thermal resistance increases nonlinearly as the power and junction temperature decrease. These features are qualitatively consistent with the predictions of a model assuming the thermal resistance is dominated by phonon radiation through an interface (computed at 20 K, red solid line). Dashed black lines are added as guides to the eye.}
\label{fig:4}
\end{figure*}

We next examine the origin of the nonlinear trend of thermal resistance versus power in Fig.~\ref{fig:4}. Assuming the total gate thermal resistance is dominated by the thermal boundary resistance, in principle a microscopic model of thermal boundary resistance could be constructed from thermal resistance versus temperature and knowledge of the phonon density of states of the semiconductor and metal. However in practice,  the gate is formed via deposition of alternating layers of Ti, Pt, Au, and subsequent bake and passivation steps induce atomic diffusion and the formation of intermetallic compounds \cite{Cha_Thesis}. As a result, knowledge of the atomic structure and vibrational modes of the interface required for such a model is lacking.

Considering these challenges, we instead construct a qualitative model for the thermal resistance in which the phonons are assumed to follow a Debye model and radiate from the gate through the interface. Assuming that the temperature is small compared to a Debye temperature associated with atomic vibrations at the interface, the heat flux through the interface obeys the Stefan-Boltzmann law $q = \epsilon \sigma_p (T_j^4 - T^4)$ \cite{Pohl_1989}, where $\epsilon$ is a measure of the transmission of phonons through the interface and $\sigma_p = \pi^{2} k_B^4/40\hbar^3 v_{ave}^2 \sim 600$ \si{\watt \per \meter \squared \per \kelvin \tothe{4}} is the Stefan-Boltzmann constant for phonons in GaAs. Here, $v_{ave} \approx 3500$ \si{\meter \per \second} is the Debye velocity in GaAs computed from the average sound velocities \cite{Blakemore_1982}. In this regime, the thermal resistance can be defined as $R_{th}^{-1} = \epsilon \sigma_p \, (T_j + T) (T_j^{2} + T^2)$.

Applying this model to the data measured at 20 K, we obtain the  curve shown in Fig.~\ref{fig:2b}. Despite the simplicity of the model, it qualitatively captures the nonlinear variation of thermal resistance with power for  $\epsilon \sim 0.02$.  We note that the thermal resistance predicted by diffusion theory based on the bulk thermal conductivity of GaAs at $T=20$ \si{\kelvin} predicts a thermal resistance of 0.5 \si{\kelvin \square \micro \meter \per \milli \watt}, orders of magnitude smaller than the observed resistance. The physical picture of heat dissipation from the gate that emerges is therefore the radiation of phonons from the gate with a heat flux that is smaller than the pure radiation value owing to phonon reflections at the gate-semiconductor interface. 

\section{Discussion}\label{sec:Discussion}

We now discuss the implications of heat dissipation by phonon radiation on the self-heating and microwave noise figure of HEMTs. In the present experiments, the gate was forward-biased while the drain was grounded so that heat was generated by current flowing through the gate. However, under the typical low noise operating conditions for depletion-mode HEMTs, the gate and drain are reverse and forward biased, respectively,  and heat is generated by current flowing through the channel. Despite these differences, the identification of the phonon radiation mechanism supported by the measurements in this study allows us to assess the magnitude of self-heating at the low-noise operating bias. Considering heat transport to occur by phonon radiation, at the low-noise bias phonons generated in the channel radiate to the gate which then radiates phonons to the substrate to balance the incoming heat flux. The steady-state temperature of the gate is set by the radiation space resistances between the gate, channel, and substrate. 

An equivalent radiation circuit model can be used to predict the temperature rise in the gate from the radiative phonon flux originating at the drain (see Eqn.~6.48 of Ref.~\cite{mills_heattransfer}). In this model, the three nodes in the circuit are the channel (c), gate (g), and substrate (s) linked by space radiation resistances. We assume all surfaces are black for simplicity. The gate node is adiabatic to excellent approximation so that all absorbed radiation is re-emitted (see Supplementary Information in Ref.~\cite{schleeh_2014}). 

Under these assumptions, the gate temperature $T_g$ can be expressed as $J_g = (J_d R_{cg} +  J R_{gs}) (R_{cg}+R_{gs})^{-1}$, where $R_{ij} = A_{i} F_{ij}$ is the space resistance between  nodes $i$ and $j$, $A_i$ is the emitting line length, $F_{ij}$ is the view factor, and $J_i = \sigma_p T_i^4$ is the blackbody emissive power from node $i$ at temperature $T_i$. For a specified substrate temperature $T_s = T$ and power density from Joule heating at the channel $Q_c =  (J_c - J_g)/R_{cg} + (J_d - J_s)/R_{cs}$, the radiation circuit can be solved for the channel and gate temperature rise. Based on typical HEMT geometry, we estimate emitting line lengths as $A_g = A_c \sim 70$ \si{\nano \meter}. The view factor is estimated from the intercept of the 2D solid angle of the gate from the emitter region in the channel. For a typical HEMT gemometry, we obtain $F_{cg} \sim 0.3$. Solving for the temperature rise, we find that the gate temperature rise is $\sim 2/3$ of that at the channel. At a typical low-noise bias, the dissipated power is $\sim 30$ \si{\milli \watt \per \milli \meter} \cite{schleeh_2014}. Taking the base temperature $T=4$ K, the temperature rise in the channel is $\sim 25$ \si{\kelvin} and the gate temperature is therefore predicted to exceed 20 K. 

The self-heating of the gate affects the microwave noise of HEMTs because the thermal noise associated with the gate resistance is added at the input. If the steady-state gate temperature exceeds the base temperature, the microwave noise will be larger than predicted based on the base temperature. The above analysis implies that the thermal noise contribution of the HEMT gate at liquid helium temperatures is several times larger than previously assumed. An incorrect gate temperature directly affects the extracted drain temperature and consequently the interpretation of the physical origin of noise in HEMTs. Finally, we note that previous work has interpreted noise saturation at liquid helium temperatures to the saturation of the drain noise added at the output (see Fig.~1d of Ref.~\cite{schleeh_2014}).  Our measurements indicate that the observed noise saturation is in fact due to elevated thermal noise as the gate temperature saturates with base temperature, as seen in Fig.~\ref{fig:3b}.

This analysis indicates that self-heating will limit the minimum noise figure of HEMT amplifiers without decreases in power consumption or improvements in device thermal management that decrease the physical temperature of the gate. Recently, low noise amplifiers with power consumption of  hundreds of \si{\micro \watt} were reported, a value that is several times lower than those of typical HEMTs \cite{Cha_2020}. While these reductions can reduce gate heating, the quartic dependence associated with phonon radiation means that even at 300 \si{\micro \watt} and $T=4$ K, the gate temperature is predicted to be  $\sim 10$ \si{\kelvin}. Therefore, additional considerations of thermal management are necessary to reduce the excess thermal noise  resulting from self-heating. A possible strategy is using direct immersion in liquid helium, which is routinely used for thermal management of superconducting magnets used in high-energy physics experiments \cite{Lebrun_1994}.

\section{Summary}

We have presented measurements of the gate junction temperature and thermal resistance of a low-noise HEMT from cryogenic to room temperature obtained using a Schottky thermometry method. The magnitude and trend of the extracted thermal resistance versus power and base temperature are consistent with heat dissipation by phonon radiation through an interface. Considering phonon radiation as the dominant mechanism of heat transfer, we estimate the intrinsic temperature of the gate at the low-noise operating bias using a radiation circuit. The model predicts that at liquid helium temperatures, the gate will self-heat to a temperature several times that of the base temperature. Our measurements thus indicate that self-heating  constitutes a practical lower limit for the minimum microwave noise figure of cryogenic HEMT amplifiers unless thermal management strategies to remove heat from the gate can be identified.

\begin{acknowledgements}
A.Y.C., B.G. and A.J.M. were supported by the National Science Foundation under Grant No. 1911220. I.E. was supported by the National Science Foundation Graduate
Research Fellowship under Grant No. DGE-1745301. Any opinions, findings, and conclusions or recommendations expressed in this material are those of the author and do not necessarily reflect the views of the National Science Foundation. J.K. and A.J.M were supported by a Jet Propulsion Laboratory PDRDF 107614-20AW0099.
\end{acknowledgements}

\bibliographystyle{is-unsrt}
\bibliography{references}{}

\end{document}